\documentclass[prl,superscriptaddress,unsortedaddress,twocolumn,%
  showpacs,preprintnumbers,amsmath,amssymb]{revtex4-1}
\usepackage{graphicx}
\usepackage[breaklinks,colorlinks=true]{hyperref}
\usepackage{MnSymbol}

\newcommand{\Nc}{N_\text{c}}

\newcommand{\bzero}{\boldsymbol{0}}

\newcommand{\bk}{\boldsymbol{k}}
\newcommand{\bp}{\boldsymbol{p}}

\newcommand{\bB}{\boldsymbol{B}}
\newcommand{\Slash}[1]{\ooalign{\hfil/\hfil\crcr$#1$}}
\newcommand{\calC}{\mathcal{C}}
\newcommand{\calE}{\mathcal{E}}
\newcommand{\calM}{\mathcal{M}}
\newcommand{\calL}{\mathcal{L}}
\newcommand{\calP}{\mathcal{P}}
\newcommand{\calS}{\mathcal{S}}
\newcommand{\calQ}{\mathcal{Q}}
\newcommand{\calT}{\mathcal{T}}
\newcommand{\calJ}{\mathcal{J}}

\newcommand{\tr}{\mathrm{tr}\,}

\newcommand{\sgn}{\mathop{\mathrm{sgn}}}
\newcommand{\feq}{f_{\text{eq}}}
\newcommand{\fbareq}{\bar{f}_{\text{eq}}}
\newcommand{\gequiv}{g_{\text{eq}}}
\newcommand{\mq}{m_q}

\begin{document}
\title{Electric conductivity of hot and dense quark matter in a magnetic field\\
       with Landau level resummation via kinetic equations}
\preprint{RIKEN-QHP-334, RIKEN-STAMP-37}
\author{Kenji Fukushima}
\affiliation{Department of Physics, The University of Tokyo, %
             7-3-1 Hongo, Bunkyo-ku, Tokyo 113-0033, Japan}
\author{Yoshimasa Hidaka}
\affiliation{Theoretical Research Division, Nishina Center, RIKEN, %
             2-1 Hirosawa, Wako, Saitama 351-0198, Japan}
\affiliation{iTHEMS Program, RIKEN, %
             2-1 Hirosawa, Wako, Saitama 351-0198, Japan}
\begin{abstract}
  We compute the electric conductivity of quark matter at finite
  temperature $T$ and quark chemical potential $\mu$ under a magnetic
  field $B$ beyond the Lowest Landau level approximation.  The
  electric conductivity transverse to $B$ is dominated by the Hall
  conductivity $\sigma_H$.  For the longitudinal conductivity
  $\sigma_\parallel$, we need to solve kinetic equations.  Then, we
  numerically find that $\sigma_\parallel$ has only mild dependence on
  $\mu$ and the quark mass $m_q$.  Moreover, $\sigma_\parallel$ first
  decreases and then linearly increases as a function of $B$, leading
  to an intermediate $B$ region which looks consistent with the
  experimental signature for the chiral magnetic effect.  We also
  point out that $\sigma_\parallel$ at nonzero $B$ remains within the
  range of the lattice-QCD estimate at $B=0$.
\end{abstract}

\pacs{25.75.-q, 25.75.Nq, 21.65.Qr, 12.38.-t}
\maketitle

\paragraph*{Introduction:}

Extreme matter of quarks and gluons in quantum chromodynamics (QCD)
could realize as a quark-gluon plasma at high energy in
nucleus-nucleus collisions and as quark matter at high baryon density
in the neutron star cores.  Nowadays, the nucleus-nucleus collision
experiment is aiming to explore the QCD phase diagram at finite
temperature $T$ and quark chemical potential $\mu$, which is called
the beam energy scan program.  Interestingly, such hot and dense QCD
matter may be exposed under a strong magnetic field $B$ if the
nucleus-nucleus collision is noncentral.  The presence of strong $B$
provides us with an ideal probe to topological contents of the QCD
vacuum, as exemplified by the chiral magnetic effect
(CME)~\cite{Fukushima:2008xe} for instance.

To quantify topological effects induced by $B$, we need to estimate
transport coefficients, among which one of most important is the
electric conductivity $\sigma$.  Indeed, the CME signature in
condensed-matter system of Weyl semimetals is the negative
magnetoregistance, that is, quadratic rise of
$\sigma_{\rm CME}(B)\propto B^2$~\cite{Son:2012bg}, which has been
first detected experimentally in Ref.~\cite{Li:2014bha} under an
assumption that nontopological $\sigma$ is insensitive to $B$.  In
contrast to condensed-matter system, for hot and dense quark matter, we
can make a first-principles estimate for $\sigma(B)$ from QCD
directly.  On top of that $\sigma(B)$ is an essential parameter for
the CME detectability in the nucleus-nucleus collision, $\sigma(B)$
controls the life time of $B$~\cite{McLerran:2013hla,Tuchin:2015oka}.

So far, $\sigma(B)$ has been perturbatively calculated in QCD under
a hierarchy of relevant scales, $\sqrt{eB} \gg T \gg gT$, where $e$
represents the charge of the proton and $g$ the QCD charge, using the
lowest Landau level approximation
(LLLA)~\cite{Hattori:2016cnt,Hattori:2016lqx}.  Usually, the LLLA is a
reasonable approximation for strong $B$ and has been adopted for
various QCD observables such as the heavy quark diffusion
constant~\cite{Fukushima:2015wck}, the bulk
viscosity~\cite{Hattori:2017qih}, etc.  The validity of the LLLA is
questionable, however, for $\sigma(B)$ involving ($u$ and $d$) quarks
with small mass $\mq$, i.e., $\sigma\to\infty$ as $\mq\to 0$ since the
scattering phase space is too severely restricted by the approximation.

In the present work we significantly revise the calculation of
$\sigma(B)$ in a different (more realistic) regime,
$\sqrt{eB} \gg gT$, which is required to justify our neglecting
scattering processes for $T$-induced quark damping ($\sim g^2 T$),
namely, $\Delta\varepsilon\simeq eB/T \gg g^2 T$ where
$\Delta\varepsilon$ is an energy gap associated with adjacent Landau
levels.  Then, we will find that our $\sigma$ with full Landau level
resummation shows much milder $\mq$ dependence than the LLLA result.
We will also see that the $B$ dependence is minor.  Thus, comparing
our finite-$B$ $\sigma$ to the lattice-QCD measured value at
$B=0$~\cite{Aarts:2007wj,Ding:2010ga,Ding:2016hua} would make sense as
a consistency check.
\vspace{0.5em}

\paragraph*{Some definitions:}

The electric conductivity is given by the following Kubo formula:
\begin{equation}
  \sigma^{ij}  = \lim_{k_0\to0} \lim_{\bk\to\bzero}
  \frac{1}{2i k_0} \bigl[ \Pi_R^{ij}(k) - \Pi_A^{ij}(k) \bigr]\,,
\end{equation}
where $\Pi_R^{\mu\nu}(k)$ and $\Pi_A^{\mu\nu}(k)$ are the retarded and
the advanced polarization functions, respectively, defined by
\begin{equation}
  \Pi_{R/A}^{ij}(k) := \pm i \int d^4 x\, e^{i k\cdot x}\,
  \theta(\pm t) \bigl\langle [j^i(x),j^j(0)] \bigr\rangle\,,
\end{equation}
where ``$+$'' is for $R$ and ``$-$'' is for $A$.  We note that, when
we work at finite density, $j^i$ in the above formula is not
necessarily the electric current,
$j_{\rm em}^i=\sum_f q_f\bar{\psi}_f \gamma^i\psi_f$ where $f$ refers
to flavor and $q_f$ is the electric charge of $f$-quark,
i.e., $q_u=(2/3)e$ and $q_d=-(1/3)e$.  The hydrodynamic mode
subtraction is needed as
$j^i = j_{\rm em}^i-n_e T^{0i}/(\calE+\calP_i)$
with the electric charge density $n_e$, the energy momentum tensor
$T^{\mu\nu}$, the energy density $\calE=\langle T^{00}\rangle$, and
the pressure $\calP_i=\langle T^{ii}\rangle$~\cite{Zubarev:1979}.

For perturbative calculations of $\Pi_{R/A}^{ij}(k)$ the free quark
propagator at finite $B$ is the essential building block.  The
retarded propagator in flavor $f$ sector is given by a sum over the
Landau levels labeled by $n$ as
\begin{equation}
  S^f_{R/A}(p) = \sum_{n=0}^\infty
  \frac{-S^f_n(p)}{p_0^2-\varepsilon_{f n}^2\pm i \epsilon p_0}
  = \sum_{n=0}^\infty
  \frac{-S^f_n(p)}{p_\parallel^2-m_{f n}^2\pm i \epsilon p_0}\,,
\end{equation}
where the (flavored) Landau quantized energy dispersion is
$\varepsilon_{f n}=\sqrt{p_z^2+2|q_f B|n+m_f^2}$ and we defined
$m_{f n}^2:=2|q_f B| n+m_f^2$, $p^\mu_\perp:=(0,p_x,p_y,0)$, and
$p^\mu_\parallel:=(p_0,0,0,p_z)$.  Here, we chose the $B$ direction
along the $z$ axis without loss of generality.  The numerator
$S^f_n(p)$ has Dirac index structures decomposed as
\begin{equation}
  \label{eq:AB}
  \begin{split}
  S^f_n(p) = (\Slash{p}_\parallel + m) \bigl[
  P^f_+ A_{n}(4\xi^f_p) + P^f_- A_{n-1}(4\xi^f_p) \bigr] \\
  + \Slash{p}_\perp B_n(4\xi^f_p)
  \end{split}
\end{equation}
with $\xi^f_p:=|\bp_\perp|^2/(2|q_f B|)$.  We introduced
$A_{n}(x) := 2e^{-x/2} (-1)^n L_n(x)$,
and
$B_n(x) := 4e^{-x/2} (-1)^{n-1} L_{n-1}^{(1)}(x)$ where
$L_n(x)=L_n^{(0)}(x)$ and
$L_n^{(\alpha)}(x)$ represent the generalized Laguerre
Polynomials~\cite{Gusynin:1994xp}.  In the above expression $P^f_\pm$
represents the projection operator,
$P^f_\pm:=(1\pm \sgn(q_f B)\,i\gamma^1\gamma^2)/2$.

We adopt the real-time Schwinger-Keldysh formalism in the $R/A$ basis
in which the standard propagators on the Schwinger-Keldysh paths (1,2)
are transformed through the following relations:
$S^f_{RA} = -i S^f_R$, $S^f_{AR} = -i S^f_A$, $S^f_{AA} = 0$, and
$S^f_{RR} = -i[1/2-n_F(p_0-\mu_f)](S_R^f-S_A^f)$, where
$n_F$ is the Fermi-Dirac distribution function.
\vspace{0.5em}

\paragraph*{Electric conductivity:}

We are now ready for proceeding to the conductivity calculation.
We decompose the anisotropic tensor structure of the
electric conductivity using $\hat{B}^i:=B^i/|\bB|$ as
\begin{equation}
  \sigma^{ij} = \sigma_H\,\epsilon^{ijk}\hat{B}^k
  + \sigma_\parallel\,\hat{B}^i \hat{B}^j
  + \sigma_\perp\,(\delta^{ij}-\hat{B}^i \hat{B}^j)\,,
\end{equation}
where $\sigma_H$ represents the Hall conductivity for an electric
current perpendicular to both electric and magnetic fields.  
In the $R/A$ basis, the polarization tensor at the one-loop order reads
\begin{align}
  \label{eq:oneloop}
  & \Pi_R^{\mu\nu}(k) = -i \sum_f q_f^2\int\frac{d^4 p}{(2\pi)^4}
  \tr\bigl[ \gamma^\mu S^f_{RR}(k+p) \gamma^\nu S^f_{AR}(p) \bigr]
  \notag \\
  & \quad -i \sum_f q_f^2 \int\frac{d^4 p}{(2\pi)^4}
  \tr\bigl[ \gamma^\mu S^f_{RA}(k+p) \gamma^\nu S^f_{RR}(p) \bigr]\,,
\end{align}
apart from the hydrodynamic mode subtraction, 
which will be taken into account later.  
We can straightforwardly perform the integration~\eqref{eq:oneloop} to
get
\begin{equation}
  \sigma_H = \frac{n_e}{B}\,,
\end{equation}
which is nothing but  the formula for the Hall conductivity.
Up to the one-loop order $\sigma_\perp=0$ which is intuitively
understood from the Landau quantization of transverse motion.  A
nonzero value of $\sigma_\perp$ appears from the two-loop and higher
order contributions.  Here, we just give a parametric estimate, that
is,
\begin{equation}
  \frac{\sigma_\perp}{T} \;\sim\; \frac{g^2T^2}{|eB|}\,,
\end{equation}
which is small in our condition of $\sqrt{|eB|}\gg gT$.  This
parametric form is derived from one self-energy insertion
to the fermion propagators.  The leading behavior of the self-energy
is $\sim g^2 T$, while the propagator is of order
$1/\Delta\varepsilon \sim T/|eB|$.  Thus, the combination of these
factors leads to $g^2 T\cdot T/|eB| = g^2 T^2/|eB|\ll1$.
\vspace{0.5em}

\paragraph*{Kinetic equations:}

\begin{figure}
  \centering
  \includegraphics[width=0.6\columnwidth]{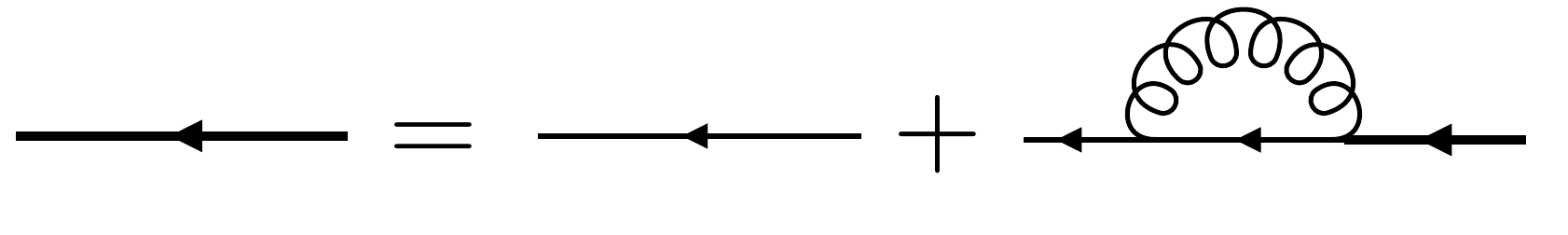} \\
  \includegraphics[width=0.5\columnwidth]{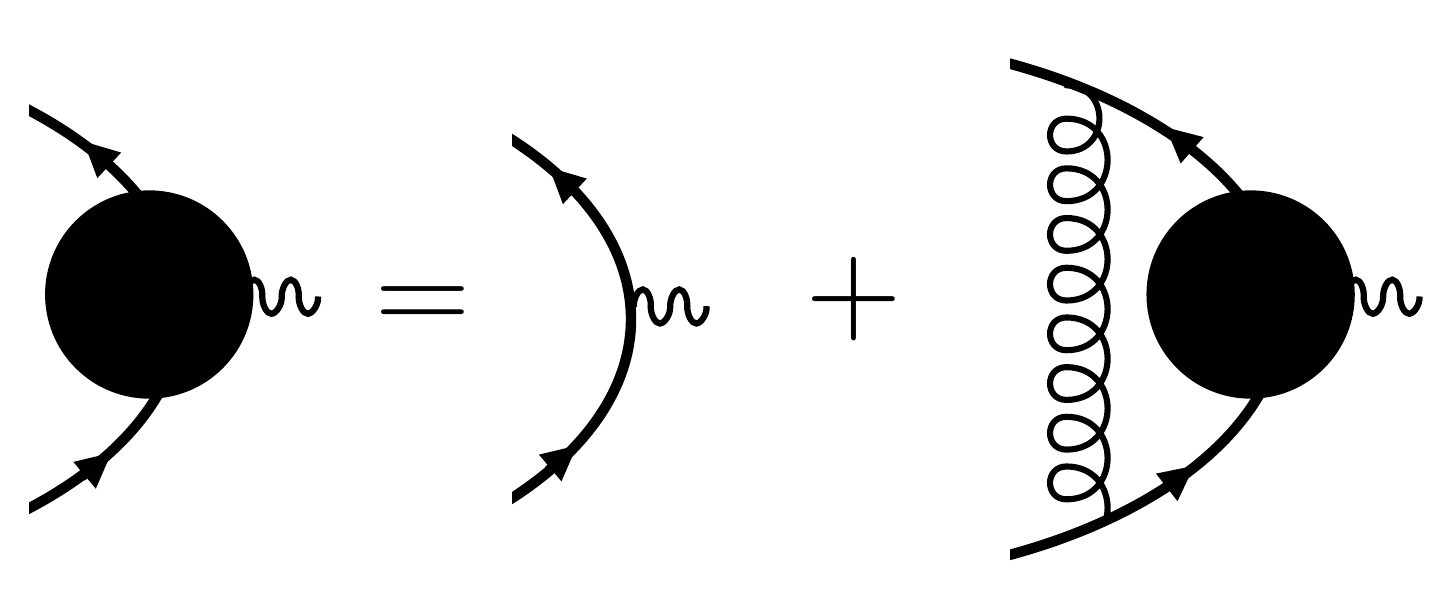}
  \caption{Illustration of the Bethe-Salpeter equations;  the resummed
  propagator with self-energy insertions (top) and the resummed
  vertex with ladder diagrams (bottom).}
  \label{fig:BSequation}
\end{figure}

Next, we calculate the longitudinal conductivity which is of our main
interest.  To this end we must deal with the resummation over pinching
singularities (see Ref.~\cite{Hidaka:2010gh} for example).  An
efficient approach to resum higher order diagrams is solving the
Bethe-Salpeter equations, as illustrated in Fig.~\ref{fig:BSequation},
which amounts to the common formalism used in
Ref.~\cite{Arnold:2000dr}.

The Bethe-Salpeter equations can be translated to the linearized
kinetic or Boltzmann equations, that is,
\begin{equation}
  \label{eq:BoltzmannEq}
  \begin{split}
  & 2P_p^\mu \bigl(\partial_\mu
  + q_f F_{\nu\mu} \partial_{p_\nu} \bigr) f_p = -C[f]\,\\
  & 2\bar{P}_{p'}^\mu \bigl(\partial_\mu
  - q_f F_{\nu\mu} \partial_{p'_\nu} \bigr) \bar{f}_{p'} = -\bar{C}[f]\,\\
  & 2k^\mu \partial_\mu g_k = -\tilde{C}[f]\,
  \end{split}
\end{equation}
for quarks, anti-quarks, and gluons, respectively, where
$\partial_{p_\nu}:=\partial/\partial p_\nu$ and $C[f]$,
$\bar{C}[f]$, and $\tilde{C}[f]$ represent the collision terms.  In
the above, $2P_p^\mu := \bar{u}(p)\gamma^\mu u(p)$ and
$2\bar{P}_{p'}^\mu := \bar{v}(p')\gamma^\mu v(p')$, with the wave
functions $u(p)$ and $v(p')$ for particle and anti-particle,
respectively, and the subscript $p$, $p'$, and $k$ represent not only
the momenta but also the Landau level $n$, the angular momentum $l$,
the spin $s$, the color $c$, and the flavor $f$ collectively.

To solve the Boltzmann equation perturbatively, we expand the
distribution functions in terms of small deviations, $\delta f_p$,
$\delta\bar{f}_p$, and $\delta g_k$, around the thermal
equilibrium, $\feq(p)=n_F(\varepsilon_{fn}-\mu)$,
$\fbareq(p)=n_F(\varepsilon_{fn}+\mu)$, and $\gequiv(k)=n_B(\omega_k)$
where $n_B$ is the Bose-Einstein distribution function and
$\omega_k=|\bk|$ is the energy of massless gluons.  It would be
more convenient to introduce $\chi_p$, $\bar{\chi}_{p'}$, and
$\tilde{\chi}_k$ rescaled by common factors as
$\delta f_p = \beta \feq(p)[1-\feq(p)]\,  E_z\, \chi_p$,
$\delta\bar{f}_{p'}=\beta\fbareq(p')[1-\fbareq(p')]\,E_z\,\bar{\chi}_{p'}$,
and $\delta g_k = \beta \gequiv(k)[1+\gequiv(k)]\,E_z\,\tilde{\chi}_k$,
where $\beta=1/T$ is the inverse temperature.

Suppose that we solved $\chi_p$, $\bar{\chi}_{p'}$, and
$\tilde{\chi}_k$ from the kinetic equations~\eqref{eq:BoltzmannEq},
we can express the electric current as
$j_z=\sigma_\parallel E_z = \sumint_p 2P_p^3 q_f (\delta f_p-\delta\bar{f}_p)$,
from which we can read $\sigma_\parallel$, where $\sumint$ denotes
the phase space sum of all quantum numbers and the invariant
integration of momentum.   In this way we come by the following
formula,
\begin{equation}
  \label{eq:cond}
  \begin{split}
  & \sigma_\parallel = \beta\Nc \sum_f \frac{q_f |q_f B|}{2\pi}
  \sum_{n=0}^\infty \alpha_n
  \int\frac{d p_z}{2\pi}\,\frac{p_z}{\varepsilon_{f n}} \\
  & \quad\times \Bigl\{\feq(p)[1\!-\!\feq(p)]\chi_p
  - \fbareq(p)[1\!-\!\fbareq(p)]\bar{\chi}_p\Bigr\}\,.
  \end{split}
\end{equation}
Here, we introduced the spin degeneracy factor $\alpha_n$ by
$\alpha_0 = 1$ and $\alpha_{n>0}=2$.

Now, let us return to our problem of solving
Eq.~\eqref{eq:BoltzmannEq}.  In the left-hand side, $\partial_0$ on
$\feq$ picks up a term $\propto \partial_0 u_z$ where $u_z$ is the $z$
component of fluid velocity which can be eliminated by the leading
order hydrodynamic equation,
$\partial_0 u_z = n_e E_z/(\calE+\calP_z)$.  Then, the left-hand side
of the first equation for quarks simplifies as
\begin{equation}
  \label{eq:BoltzmannEq2}
  2P_p^0 \,\bigl(\partial_0 + q_f E_z\,\partial_{p_z} \bigr) f_p
  = -\beta W_p E_z
  \biggl( q_f \frac{p_z}{\varepsilon_{f n}} - \frac{n_e p_z}
  {\calE + \calP_z} \biggr)\,.
\end{equation}
Here, we defined $W_p:=2P_p^0\, \feq(p)[1-\feq(p)]$.  The second
kinetic equation for $\bar{f}_p$ has the same structure as above with
$f_p$, $W_p$, and $q_f$ replaced with $\bar{f}_p$,
$\bar{W}_p:=2P_p^0\, \fbareq(p)[1-\fbareq(p)]$, and $-q_f$.  Likewise,
the gluon equation is $2\omega_k \,\partial_0 g_k
= -\beta \tilde{W}_k (-k_z\partial_0 u_z)$ with
$\tilde{W}_k:=2\omega_k\, \gequiv(k)[1+\gequiv(k)]$.

Using the following multi-component symbols,
\begin{equation}
  \label{eq:J}
  \calJ^\mu := q_f\begin{pmatrix}
  p^\mu / \varepsilon_{f n} \\
  -p'^\mu / \varepsilon_{f n'} \\
  0
  \end{pmatrix}\,,\qquad
  \calT^{0\mu} := \begin{pmatrix}
  p^\mu \\ p'^\mu \\ k^\mu
  \end{pmatrix}\,,
\end{equation}
we can summarize three kinetic equations as
\begin{equation}
  \label{eq:BoltzmannSymbol}
  \calS:= \calJ^z-\frac{n_e \calT^{0z}}{\calE+\calP_z} = \calL \chi\,,
\end{equation}
where the left-hand side will be denoted by $\calS$ in what follows,
and the right-hand side represents the collision terms;  $\calL$
is a linear operator defined by
\begin{equation}
  \calL \chi := \calL \begin{pmatrix} \chi_p \\
  \bar{\chi}_{p'} \\ \tilde{\chi}_k
  \end{pmatrix} = \frac{1}{\beta E_z}
  \begin{pmatrix}
   C[f] / W_p \\
   \bar{C}[f] / \bar{W}_{p'} \\
   \tilde{C}[f] / \tilde{W}_k
   \end{pmatrix}\,.
\end{equation}
We should then solve $\chi=\calL^{-1}\calS$ using our symbolic
notation.  We note that $\calL$ contains five zero eigenvalues (for a
single flavor and more for multi flavors) with the eigenvectors,
$\calC^a = \{\calJ^0 ,\calT^{0\mu} \}$,
corresponding to the charge and the energy-momentum conservations.
For two flavors $\calC^a$ also contains the quark number
conservation.

To formulate the projection procedure, let us introduce an inner
product for two functions,
$A=(a_p,\bar{a}_{p'},\tilde{a}_{k})$ and
$B=(b_p,\bar{b}_{p'},\tilde{b}_{k})$, as follows
\begin{equation}
  \label{eq:inner}
  (A, B) := \int_p\, W_p\, a_p b_p
  + \int_{p'} \bar{W}_{p'}\, \bar{a}_{p'} \bar{b}_{p'}
  + \int_k\, \tilde{W}_k\, \tilde{a}_k \tilde{b}_k \,.
\end{equation}
It is then easy to rewrite Eq.~\eqref{eq:cond} as
$\sigma_\parallel =\beta(\calJ^z, \chi)$ using Eq.~\eqref{eq:J}.  Now,
with the zero eigenvectors $\calC$ and the inner product, we define a
projection operator onto functional space excluding zero eigenvalues
as
\begin{equation}
  \calQ O := O - \sum_{a,b} \calC^a (\calC,\calC)^{-1}_{ab}
  (\calC^b, O) \,,
\end{equation}
where $(\calC,\calC)^{-1}_{ab}$ is the inverse matrix of
$(\calC^a,\calC^b)$.  We see $\calQ^2 = \calQ$ and $\calQ\calC^a=0$ by
construction.  Using alternative expressions for the charge density
and the enthalpy, i.e.,
$n_e = \beta(\calT^{0z}, \calJ^z)$ and
$\calE+\calP_z = \beta(\calT^{0z},\calT^{0z})$~\cite{Minami:2012hs}, we can write
$\calS = \calQ\calJ^z$.  Noting $\calL=\calL\calQ$, we find the formal
solution of $\calL\chi=\calS$ as $\chi = \calQ\calL^{-1}\calQ\calS$,
where $\calQ\calL^{-1}\calQ$ satisfies a relation
$\calL\calQ\calL^{-1}\calQ=\calQ$.  We eventually obtain
\begin{equation}
  \label{eq:sigmazz}
  \sigma_\parallel = \beta(\calJ^z, \calQ\calL^{-1}\calQ\calS)
  = \beta(\calS, \calL^{-1} \calS)\,.
\end{equation}
This means, the zero modes of $\calL$ are already projected out once
applied on $\calS$.
\vspace{0.5em}

\paragraph*{Collision terms:}

\begin{figure*}
  \centering
  \begin{minipage}{0.2\columnwidth}
    \includegraphics[width=\textwidth]{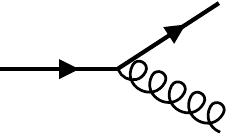}\\
    (a1) $q\to qg$
  \end{minipage} \hspace{2em}
  \begin{minipage}{0.2\columnwidth}
    \includegraphics[width=\textwidth]{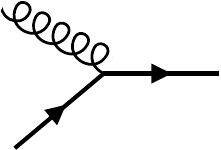}\\
   (a2) $qg\to q$
  \end{minipage} \hspace{2em}
  \begin{minipage}{0.2\columnwidth}
    \includegraphics[width=\textwidth]{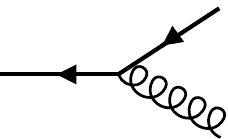}\\
   (b1) $\bar{q}\to \bar{q}g$
  \end{minipage} \hspace{2em}
  \begin{minipage}{0.2\columnwidth}
    \includegraphics[width=\textwidth]{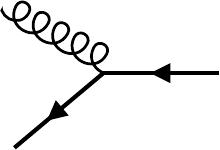}\\
   (b2) $\bar{q}g\to \bar{q}$
  \end{minipage} \hspace{2em}
  \begin{minipage}{0.2\columnwidth}
    \includegraphics[width=\textwidth]{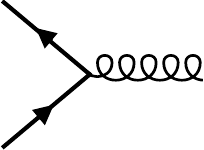}\\
   (c1) $q\bar{q}\to g$
  \end{minipage} \hspace{2em}
  \begin{minipage}{0.2\columnwidth}
    \includegraphics[width=\textwidth]{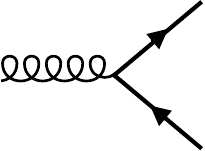}\\
   (c2) $g\to q\bar{q}$
  \end{minipage}
  \caption{Diagrams of the synchrotron radiation process with a quark
    (a1), with an anti-quark (b1), and the pair annihilation (c1).
    Their inverse processes are (a2), (b2), and (c2), respectively.}
  \label{fig:rad}
\end{figure*}

The collision term is the most complicated part of our calculations.
For $\sqrt{eB} \gg gT$, the $1\leftrightarrow 2$ process of typical
scale $\sim g^2 eB/T^2$ is dominant as compared to the
$2\leftrightarrow 2$ process of typical scale $\sim g^4$.  For the
$1\leftrightarrow 2$ process there are three distinct contributions,
\begin{equation}
  C[f] = C_{q\to qg}[f] + C_{qg\to q}[f] + C_{q\bar{q}\to g}[f]\,,
\end{equation}
where the subscripts represent processes illustrated in
Fig.~\ref{fig:rad}.  We can also consider similar decompositions for
$\bar{C}$ for anti-quarks and $\tilde{C}$ for gluons.

The collision terms take a standard expression in terms of
distribution functions and the scattering amplitude.  After tedious
calculations we find that the scattering amplitudes of the synchrotron
radiation and the pair annihilation processes,
$i \calM_{p\to k+p'} = i g\, \bar{u}(p')\gamma^\mu t_a u(p)
  \varepsilon^\ast_\mu(k)$ and
$i \calM_{p+p'\to k} = i g\, \bar{v}(p')\gamma^\mu t_a u(p)
  \varepsilon^\ast_\mu(k)$,
can be squared with the summation over the quantum numbers and the
phase space, leading to
\begin{align}
  \label{eq:MX}
  & \int_{k,p,p'} |\calM_{p\to p'+k}|^2
  (2\pi)^4\delta^{(4)}(k - p + p') \notag\\
  & = -\frac{1}{2} \sum_{f,\,n>n'} \int\frac{d p_z}{2\pi}
  \frac{1}{2\varepsilon_{f n}}
  \int_{p_{z-}'}^{p_{z+}'} \frac{d p_z'}{2\pi}\frac{1}{2\varepsilon_{f n'}}
  X(n,n',\xi_-^f), \\
  \label{eq:MX2}
  & \int_{k,p,p'} |\calM_{p+p'\to k}|^2
  (2\pi)^4\delta^{(4)}(p + p' - k) \notag\\
  & = \frac{1}{2} \sum_{f,\,n,n'} \int\frac{d p_z}{2\pi} \frac{1}{2\varepsilon_{f n}}
  \int\frac{d p_z'}{2\pi} \frac{1}{2\varepsilon_{f n'}} X(n,n',\xi_+^f),
\end{align}
where the allowed range of $p_z'$ is restricted for the
synchrotron radiation in Eq.~\eqref{eq:MX} as
$p_{z-}' < p_z' < p_{z+}'$ with
\begin{equation}
  p_{z\pm}' = p_z \frac{m_{f n}^2 + m_{f n'}^2}{2m_{f n}^2}
  \pm \frac{m_{f n}^2-m_{f n'}^2}{2m_{f n}^2}\sqrt{m_{f n}^2+p_z^2}\,.
\end{equation}
Two integrands in Eqs.~\eqref{eq:MX} and \eqref{eq:MX2} are identical,
i.e., $X(n,n',\xi_k) := g^2 \Nc C_F \int\frac{d^2 p_\perp}{(2\pi)^2}
\tr\bigl[ \gamma_\mu S^f_n(p) \gamma^\mu S^f_{n'}(p-k) \bigr]$ with a
group factor $C_F:=(\Nc^2-1)/(2\Nc)$, except for the kinematical
constraint, that is, the argument of $X(n,n',\xi_\pm^f)$ is given by
\begin{equation}
  \label{eq:xi}
  \xi^f_\pm = \frac{(\varepsilon_{f n} \pm \varepsilon_{f n'})^2
    - (p_z \pm p_z')^2}{2|q_f B|}\,.
\end{equation}
Using Eq.~\eqref{eq:AB} and properties of the Laguerre polynomials we
find
\begin{align}
  \label{eq:X}
  & X(n,n',\xi) = g^2\Nc C_F \frac{|q_f B|}{2\pi}\,
  e^{-\xi}\frac{n!}{n'!}\, \xi^{n'-n} \biggl\{ \Bigl[ 4m_f^2 \notag\\
  &\quad -4|q_f B|(n+n'-\xi)\frac{1}{\xi}(n+n')\Bigr] F(n,n',\xi) \notag\\
  &\quad + 16|q_f B|n'(n+n')\frac{1}{\xi}
  L_n^{(n'-n)}(\xi) L_{n-1}^{(n'-n)}(\xi)\biggr\} \,,\\
  & F(n,n',\xi) :=
  \begin{cases}
    \displaystyle 1 & (n=0) \\
    \displaystyle \bigl[ L_n^{(n'-n)}(\xi) \bigr]^2
    + \frac{n'}{n} \bigl[ L_{n-1}^{(n'-n)}(\xi) \bigr]^2
    & (n>0) \,.
  \end{cases}
\end{align}

\paragraph*{Recovery of the lowest Landau level approximation:}

It would be an instructive check to see that the LLLA result is
correctly recovered in the limit of $e B\gg T^2$ (at $\mu=0$).  Since
the synchrotron radiation changes the Landau level, we can safely
discard it.  For the pair annihilation process, $X(n=0,n'=0,\xi)$
given in Eq.~\eqref{eq:X} simplifies as
$X(0,0,\xi_+^f) = 4m_f^2 g^2\Nc C_F \frac{|q_f B|}{2\pi} e^{-\xi_+^0}$
with
$\xi_+^0=[(\sqrt{p_z^2+m_f^2}+\sqrt{p_z'^2+m_f^2})^2-(p_z+p_z')^2]/(2|q_f B|)$
which is nothing but $\xi_+$ in Eq.~\eqref{eq:xi} with $n=n'=0$.
When $|q_{f}B|$ is much larger than any other scales, we can
approximate $e^{-\xi_+^0} \approx 1$.  Then, the linearized kinetic
equations reduce to a simple form as
\begin{equation}
\begin{split}
  & q_f \Nc \frac{|q_f B|}{2\pi}\beta \feq(p)[1-\feq(p)]
  \frac{p_z}{\varepsilon_{f 0}} = 4m_f^2 g^2\Nc C_F \\
  &\times \beta \frac{|q_f B|}{2\pi}
  \cdot\frac{1}{4\varepsilon_{f 0}}
  \int\frac{d p'_z}{2\pi}\frac{1}{2\varepsilon_{f 0}'}
  \feq(p) \fbareq(p')[1\!+\!\gequiv(k)] \chi_p\,,
  \end{split}
\end{equation}
where $\varepsilon_{f 0}=\sqrt{p_z^2+m_f^2}$ and
$\varepsilon_{f 0}'=\sqrt{p_z'^2+m_f^2}$.  Here we do not have to
consider mixing terms with $\bar{\chi}_{p'}$.  In this special limit,
$\calL$ is not really a matrix and we do not need to take its matrix
inversion.  Actually, we can easily solve the above kinetic equation
to obtain $\chi_p$.  Thanks to the charge conjugation symmetry, the
solution for anti-quarks is $\bar{\chi}_p=-\chi_p$.  Summarizing them,
we finally arrive at the LLLA result from Eq.~\eqref{eq:sigmazz} as
\begin{equation}
  \label{eq:LLLA}
  \begin{split}
    \sigma_\parallel &= \sum_f\frac{\Nc \beta}{g^2 C_F m_f^2} 
    q_f^2\frac{|q_f B|}{2\pi}
    \int\frac{d p_z}{2\pi}\frac{p_z^2}{\varepsilon_{f 0}} \\
    &\qquad\qquad\qquad\times \frac{\feq(p)[1-\feq(p)]^2}
    {\displaystyle \int\frac{d p'_z}{2\pi}
     \frac{1}{\varepsilon_{f 0}'}\fbareq(p')[1+\gequiv(k)]} \,,
  \end{split}
\end{equation}
which is consistent with Ref.~\cite{Hattori:2016lqx}.
\vspace{0.5em}

\paragraph*{Numerical results and discussions:}

Now we have all necessary ingredients to write down the matrix
elements of $\calL$ as a phase space convolution of
$X(n,n',\xi_\pm^f)$ and the distribution functions, $\feq$, $\fbareq$,
and $\gequiv$.  Besides the flavor $f$ and the Landau level $n$, we
should choose the complete set basis for functions of $p_{z}$, $k_{z}$ and $k_{\perp}$,
which we will take the simplest polynomial form as
$\hat{p}_z|p_z|^m$ for (anti-)quarks and $k_\perp^m$ for gluons with
integral $m$.

\begin{figure}
  \centering
  \includegraphics[width=0.8\columnwidth]{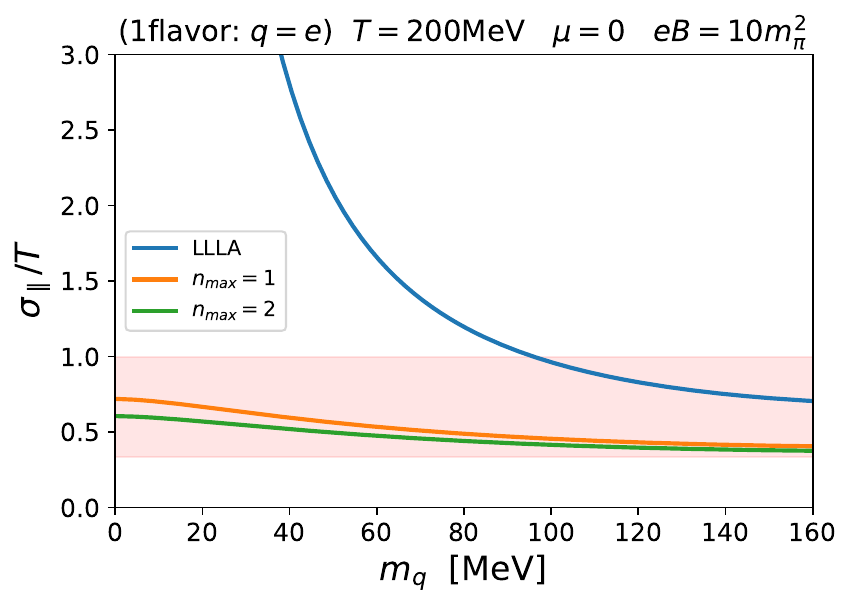}
  \caption{Mass dependence of $\sigma_\parallel$ for single flavor at
    $T=200\,\text{MeV}$, $\mu=0$, $eB=10m_\pi^2$, and $g^{2}/(4\pi)=0.3$.  The shaded
    region is the lattice-QCD estimate from Ref.~\cite{Ding:2010ga}.}
  \label{fig:massDep}
\end{figure}

Figure~\ref{fig:massDep} shows our numerical results for the quark
mass dependence of $\sigma_\parallel/T$ for a fictitious single flavor with
$q=e$ at finite $T$ and $B$ but at zero $\mu$.  We choose that the 
QCD charge as $g^{2}/(4\pi)=0.3$. We clearly see that the LLLA has 
artificial enhancement as $\mq$ approaches zero.  For
the numerical calculation we truncate the Landau level at
$n_{\rm max}$.  In the $eB=10m_\pi^2$ case, the convergence of the
Landau level sum is very fast and $n_{\rm max}=1$ already gives a good
approximation, even though the LLLA badly breaks down in the small
$\mq$ region.  It is interesting that our result is quantitatively
consistent with the lattice-QCD estimate,
$0.3\le \sigma/T\le 1.0$ (for the quark charge squared sum
$C_{\rm em}=1$)~\cite{Ding:2010ga}, which is indicated by the shaded
region in Fig.~\ref{fig:massDep}.

\begin{figure}
  \centering
  \includegraphics[width=0.8\columnwidth]{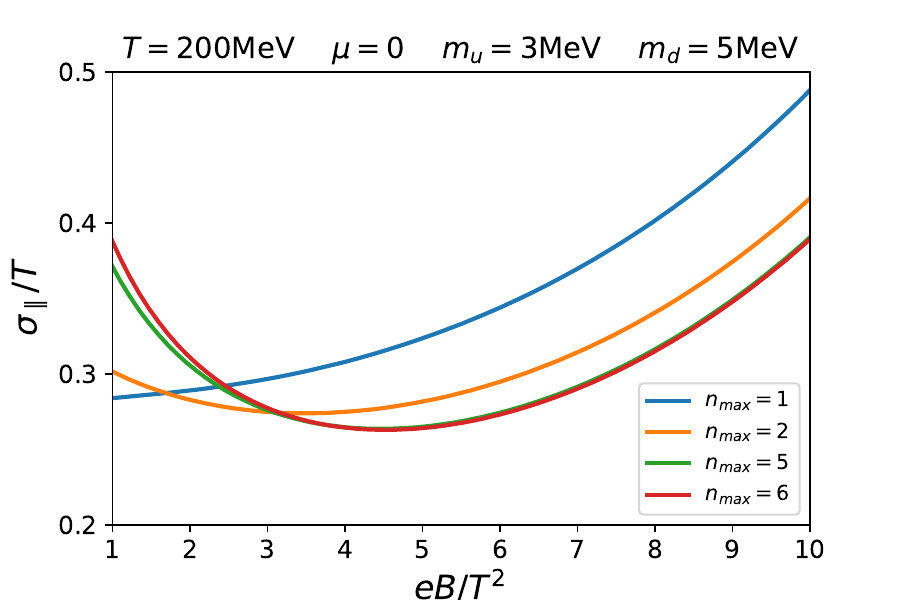}
  \caption{Magnetic and $n_{\rm max}$ dependence of
    $\sigma_\parallel/T$.}
  \label{fig:BDep}
\end{figure}

The $B$ dependence of $\sigma_\parallel/T$ has a nonmonotonic
structure as shown in Fig.~\ref{fig:BDep}, for which we adopted a
physical parameter set with $u$ and $d$ quarks.  For small
$n_{\rm max}$ or strong $B$, the LLL contribution is dominant, and
then $\sigma_\parallel$ is linearly proportional to $B$ (reflecting
the fact that the charge carrier increases), which explains the
growing behavior at large $B$ in Fig.~\ref{fig:BDep}.  When $B$ is not
such large, contributions from higher Landau levels lead to a larger
interaction cross section due to the phase space factor, which pushes
$\sigma_\parallel$ down with larger $B$.  As a result of the interplay
of these competing effects, in an intermediate region of $B$, the
increasing behavior of $\sigma_\parallel$ looks quadratic;  moreover,
this nonmonotonic behavior is consistent with what is seen in the CME
experiment in Ref.~\cite{Li:2014bha}.  Although quantitative details
may depend on underlying theory, qualitative features should be
the same for general physical systems (but could be different with
different approximations, say, the relaxation time approximation for
the collision term~\cite{Son:2012bg} may lead to a different $B$
dependence).

\begin{figure}
  \centering
  \includegraphics[width=0.8\columnwidth]{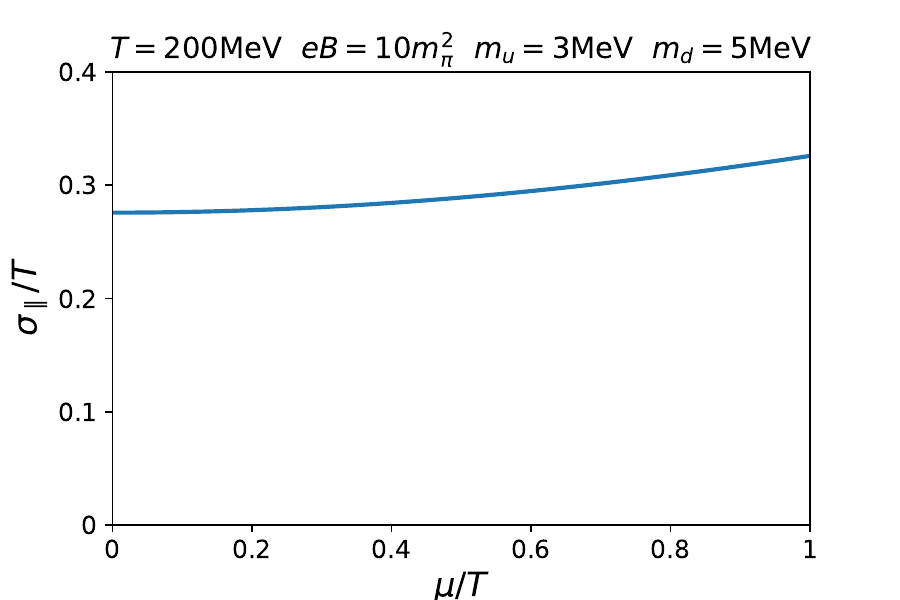}
  \caption{Density dependence for $n_\text{max}=2$.}
  \label{fig:muDep}
\end{figure}

Finally, we discuss the $\mu$ dependence as shown in
Fig.~\ref{fig:muDep}.  It is surprising at a first glance that
$\sigma_\parallel$ is rather insensitive to $\mu$.  This can be
qualitatively understood from the fact that the carrier density is
different from the net particle number but is related to the sum of
particle and anti-particle numbers.  This latter quantity is not much
changed by $\mu$ which causes imbalance between particles and
anti-particles.  In the future our estimated $B$ dependence of
$\sigma_\parallel$ could be tested by the lattice-QCD simulation at
finite $B$, while our calculation at finite $\mu$ would be a unique
prediction.  The full details of the analytical derivations and the
numerical procedures will be provided in the follow-up paper.

\begin{acknowledgments}
The authors thank
Koichi~Hattori,
Daisuke~Satow,
and
Misha~Stephanov
for useful comments and discussions.
This work was supported by Japan Society for the Promotion of Science
(JSPS) KAKENHI Grant No.\ 15H03652, 15K13479, and 16K17716.
\end{acknowledgments}

\bibliography{magnetic}
\bibliographystyle{apsrev4-1}

\end{document}